\documentclass[12pt,reqno]{amsart}
\usepackage{latexsym}

\newcommand{\K}{{\mathbf K}}
\newcommand{\tr}{{\operatorname{Tr}}}

\newcommand{\crit}{{\operatorname {crit}}}

\newcommand{\kahler}{K\"ahler }

\newcommand{\R}{{\mathbb R}}

\newcommand{\vol}{{\operatorname{Vol}}}

\newcommand{\ical}{\mathcal{I}}

\newcommand{\ncal}{\mathcal{N}}

\newcommand{\rcal}{\mathcal{R}}
\newcommand{\al}{\alpha}
\newcommand{\be}{\beta}
\newcommand{\ga}{\gamma}
\newcommand{\Ga}{\Gamma}
\newcommand{\la}{\lambda}

\newcommand{\de}{\delta}

\newcommand{\om}{\omega}

\newtheorem{theo}{{\sc Theorem}}[section]
\newtheorem{cor}[theo]{{\sc Corollary}}
\newtheorem{lem}[theo]{{\sc Lemma}}

\newenvironment{defin}{\medskip\noindent{\it Definition:\/} }{\medskip}
\newenvironment{acknowledgements}{\bigskip\noindent{\it Acknowledgements:\/} }{\medskip}

\title[Metric Dependence and Asymptotic Minimization of Critical Points] {Metric Dependence and Asymptotic Minimization of the Expected Number of Critical Points of Random Holomorphic Sections}
\author{Benjamin Baugher}
\address{Department of Mathematics, Johns Hopkins University, Baltimore, MD 21218, USA} 
\email{bbaugher@math.jhu.edu}
\date{Mar. 19, 2007}

\begin{document}

\date{Nov. 7, 2007}

\begin{abstract}

We prove the main conjecture from \cite{DSZ2} concerning the metric dependence and asymptotic minimization of the  expected number $\ncal^\crit_{N,h}$ of critical points of random holomorphic sections of the $N$th tensor power of a positive line bundle.  The first non-topological term in the asymptotic expansion of 
$\ncal^\crit_{N,h}$ is the the Calabi functional multiplied by the constant $\be_2(m)$ which depends only on the dimension of the manifold.  We prove that $\be_2(m)$ is strictly positive in all dimensions, showing that the expansion is non-topological for all $m$, and that the Calabi extremal metric, when it exists, asymptotically minimizes $\ncal^\crit_{N,h}$.

\end{abstract}

\maketitle

\section{ Introduction}\label{A1}

In the three articles \cite{DSZ1,DSZ2,DSZ3}, M. Douglas, B. Shiffman, and S. Zelditch presented their research into the statistics of critical points of random holomorphic sections of  line bundles over complex manifolds.  This line of research was initially motivated by the vacuum selection problem in string theory, and thus there was much discussion in these papers about the application of these statistics to string theory and related fields.  However, this research is also of independent interest from a geometric point of view.  The purpose of the present article is to prove the main conjecture in \cite{DSZ2} which concerns the purely geometric consideration of determining the metric dependence and asymptotic minimization of the expected number of critical points of random sections of positive line bundles. 

While it is clear that the number of critical points of a given holomorphic section will vary with the metric, it is not clear whether the same is true for the expected number of critical points of a random section.  Therefore, in \cite{DSZ2}, the authors performed an asymptotic analysis 
of the expected number $\ncal^\crit_{N,h}$ of critical points of random holomorphic sections of $(L^N,h^N) \to M^m$ in order to determine the metric dependence of this statistic.  They showed that the asymptotic expansion of $\ncal^\crit_{N,h}$ is topologically invariant to at least two orders in $N$, but that the third term in the expansion is a sum of a topological invariant and a universal constant, $\be_2(m)$, times the Calabi functional $\int_M\rho_h^2d\vol_h$, where $\rho_h$ is the scalar curvature of the \kahler form $\omega_h = \frac{i}{2} \Theta_{h}$.  The authors derived a complicated integral formula for $\be_2(m)$ and, based on their calculations in low dimensions and a heuristic argument, conjectured that $\be_2(m)>0$ for all $m$.  In this paper we will prove their conjecture.

The fact that $\be_2(m)$ does not vanish proves that the expansion is non-topological, with a metric dependence in the third term for all $m$.  The fact that $\be_2(m)$ is positive proves that, whenever it exists, the Calabi extremal metric asymptotically minimizes $\ncal^\crit_{N,h}\,$.  It would seem likely that metrics which minimize the expected number of critical points would be ideally suited for the vacuum selection problem.

\subsection{Background and Notation}\label{A2}

In this paper we consider the $N$th tensor power of a positive Hermitian line bundle $(L^N,h^N) \rightarrow (M^m,\omega_h)$ over a compact \kahler manifold of dimension $m$.  The \kahler form is given by $\omega_h = \frac{i}{2} \Theta_{h}$, where $\Theta_{h} = - \partial \bar \partial \log h$ is the curvature form of the metric on the bundle.  For a holomorphic section $s \in H^0(M,L^N)$, we consider the connection critical points given by $\nabla s(z)=0$, where $\nabla$ is chosen to be the Chern connection of $h^N$.  We will use $Crit(s,h^N)$ to denote the set of critical points of $s$. 

In general, the critical point equation given above is not holomorphic.  In a local frame $e$ we can write $s=fe$ and then
$ \nabla s = \left( \partial f - f \partial K \right) \otimes e_L$, where $K = - \log \| e \|_{h^N}^2$ is the \kahler potential.
From this we see that the critical point equation in the local frame is $\partial f - f \partial K = 0$, which is holomorphic only when $K$ is.  Therefore the cardinality of $Crit(s,h^N)$ is a non-constant random variable on the space $H^0(M,L^N)$.

Next, we endow the space $H^0(M,L^N)$ with the Gaussian measure $\gamma_{N}$ given by \begin{equation*}d\gamma_{N}(s)=\frac 1
{\pi^d}e^{-\|c\|^2} dc\;,\qquad  s=\sum _{j=1}^d
c_je_j,\end{equation*} where $dc$ is Lebesgue measure and $\{e_j\}$
is an orthonormal basis of $H^0(M,L^N)$ relative to the inner product \begin{equation*} \langle s_1, s_2 \rangle = \frac{1}{m!} \int_M h^N(s_1(z), s_2(z)) \ \omega_h^m \end{equation*} induced by $h^N$ on $H^0(M,L^N)$. 
We define the expected distribution of critical points of $s \in
H^0(M, L^N)$ with respect to  $\gamma_{N}$ to be
\begin{equation*}  \K^\crit_{N,h}  =\int_{H^0(M, L)}
 \bigg[\,\sum_{z\in Crit (s, \, h^N)} \delta_{z}\bigg]\,d\ga_N(s),
\end{equation*} 
where $\delta_{z}$ is the Dirac point mass at $z$, and then the expected number of critical points is given by \begin{equation*}\ncal^\crit_{N,h}=\int_M \K^\crit_{N,h}(z).
\end{equation*}

We recall that the Morse index $q$ of a critical point of a real-valued function is given by the number of negative eigenvalues of its Hessian at the point, and that for a positive line bundle $m \leq q \leq 2m$ \cite{Bo}.  Since the critical points of $s$ with respect to $\nabla$ are the same as those of the real-valued function $\log \|s\|_{h^N}^2$, we can consider the Morse indices of the critical points in $Crit(s,h^N)$.  To this end, we let $\ncal^\crit_{N,q,h}$ denote the expected number of critical points of Morse index $q$.  It follows that
\begin{equation*} 
\ncal^\crit_{N,h}=\sum_{q=m}^{2m}\ncal^\crit_{N,q,h}\;.\end{equation*}

In \cite{DSZ2}, the authors derived the complete asymptotic expansion of $\ncal^\crit_{N,q,h}$.

\begin{theo} \label{DSZTheo1}  Let $(L,h)\to (M,\om_h)$ be a
positive holomorphic line bundle on a compact \kahler manifold,
with $\om_h= \frac i2 \Theta_h$, then the expected  number of
critical points of Morse index $q$ ($m\le q\le 2m$) of random
sections in $H^0(M,L^N)$ has the asymptotic expansion
\begin{eqnarray}\ncal^\crit_{N,q,h} &\sim
&\left[\frac{\pi^m b_{0q}}{m!}\,c_1(L)^m\right]N^m +
\left[\frac{\pi^m\be_{1q}}{(m-1)!}\, c_1(M)\cdot
c_1(L)^{m-1}\right]N^{m-1} \nonumber
\\& & \ + \biggl[\be_{2q}\int_M\rho_h^2d\vol_h + \be'_{2q}\, c_1(M)^2\cdot
c_1(L)^{m-2} + \be''_{2q}\, c_2(M)\cdot c_1(L)^{m-2}\biggr]
N^{m-2}+\cdots\,,\nonumber\end{eqnarray} where
$b_{0q},\be_{1q},\be_{2q},\be'_{2q},\be''_{2q}$ are universal constants
depending only on the dimension $m$, and $\rho_h$ is the scalar curvature of $\om_h$.
\end{theo}

\noindent For the sake of clarity we state the obvious corollary:

\begin{cor} \label{DSZTheo2}  Under the same conditions as above, the expected  number of
critical points of random
sections in $H^0(M,L^N)$ has the asymptotic expansion
\begin{eqnarray}\ncal^\crit_{N,h} &\sim
&\left[\frac{\pi^m b_{0}}{m!}\,c_1(L)^m\right]N^m +
\left[\frac{\pi^m\be_{1}}{(m-1)!}\, c_1(M)\cdot
c_1(L)^{m-1}\right]N^{m-1} \nonumber
\\& & \ + \biggl[\be_{2}\int_M\rho_h^2d\vol_h + \be'_{2}\, c_1(M)^2\cdot
c_1(L)^{m-2}  + \be''_{2}\, c_2(M)\cdot c_1(L)^{m-2}\biggr]
N^{m-2}+\cdots\,,\nonumber\end{eqnarray} where
$b_{0},\be_{1},\be_{2},\be'_{2},\be''_{2}$ denote the sum over $q$ of $b_{0q},\be_{1q},\be_{2q},\be'_{2q},\be''_{2q}$, respectively.
\end{cor}

In our previous paper (\cite{Ba}) we examined the constant $b_0$ in the leading term of the above expansion.  We started with an integral formula for $b_{0q}$, which was derived in \cite{DSZ2}, and  simplified it using an argument based on the Itzykson-Zuber integral formula.  We then used a variant of the Selberg integral formula to obtain an exact formula for the $q=m$ case and, finally, showed that the value of the integral decreased as $q$ increased using a change of variable argument.  This gave us upper and lower bounds on the leading term of the expansion of $\ncal^\crit_{N,h}$.
\begin{theo} \label{Conj}
Let $n_q(m):=\frac{\pi^m}{m!}\,b_{0q}(m)$ denote the
leading coefficient in the expansion of $\ncal^\crit_{N,q,h}$, and let $n(m)=\sum_{q=m}^{2m} n_q(m)$, so that $$\ncal^\crit_{N,q,h}
\sim n_q(m)\,c_1(L)^m\,N^m \qquad \text{and} \qquad \ncal^\crit_{N,h}
\sim n(m)\,c_1(L)^m\,N^m .$$  Then $$n_m(m)=
2\,\frac{m+1}{m+2} \qquad \text{and} \qquad n_{q+1}(m) <
 \left(\frac{2m-q}{2m-q+1}\right)^2 n_{q}(m)\;,$$ and hence 
$$2\,\frac{m+1}{m+2} <n(m) <
\frac{2m+3}{3}\;.$$
\end{theo}
\noindent Using similar techniques, we also derived the growth rate of $\ncal^\crit_{N,h}$ as the dimension of the manifold gets large in the special case of positive line bundles over complex projective space.  

In this paper we will focus our attention on the constant $\be_{2}(m):=\sum_{q=m}^{2m}\be_{2q}(m)$.  We see from Corollary \ref{DSZTheo2} that the first two terms in the expansion are topological and that the only non-topological part of the third term is $\be_{2}$ times the Calabi functional $\int_M\rho_h^2d\vol_h$.  Hence, in order to determine the metric dependence of the expansion we will need to evaluate $\be_{2}$.  

As the analysis is simplified by treating the contributions from critical points of different Morse indices separately, the following integral formula for $\be_{2q}$ was derived in \cite{DSZ2}.  

\begin{lem} \label{L1}
\begin{equation}\label{A}\be_{2q}(m)= \frac {(-i)^{m(m-1)/2}} {4\,\pi^{2m}\prod_{j=1}^{m-1} j!}
\int_{Y_{2m-q}}\int_\R\cdots\int_\R\Delta(\la)\,\Delta(\xi)\,
\prod_{j=1}^m |\la_j|\;e^{i\langle
\la,\xi\rangle}\,\ical(\la,\xi)\, \,d\xi_1\cdots d\xi_m\,
d\la\;,\end{equation} where \begin{equation*} \ical(\la,\xi)=
\frac{F(D(\la))+\left[ \frac {4\sum_{j=1}^m \la_j} {m(m+1)(m+3)}
-\frac{2}{m+1}\right]\frac 1{\left(1 - \frac i2
\sum_j\xi_j\right)}+ \frac 2{(m+1)(m+3)\left(1 - \frac i2
\sum_j\xi_j\right)^2}} {\left(1 - \frac i2 \sum_j\xi_j\right)\prod_{j\le k}\left[ 1
+\frac i2 (\xi_j+\xi_k)\right] } \;.
\end{equation*}
Here, $Y_p=\{\la\in\R^m: \la_1>\cdots >\la_p>0>\la_{p+1}>\cdots
>\la_m\}\;$, $\Delta(\lambda) =
\Pi_{i < j} (\lambda_i - \lambda_j)$ is the Vandermonde
determinant, $D(\lambda)$ is the diagonal matrix with diagonal entries
$\lambda = (\lambda_1, \dots, \lambda_m)$, and \begin{equation}\label{F(P)} F(P)= 1 - \frac {4\, \tr\, P }{m(m+1)}+\frac
{4(\tr\,P)^2+8\,\tr(P^2)} {m(m+1)(m+2)(m+3)}\;,\end{equation}for (Hermitian)
$m\times m$ matrices $P$.  The iterated $d\xi_j$ integrals are defined in the
distribution sense.
\end{lem}

We ask the interested reader to refer to \cite{DSZ2} for additional background information.

\subsection{Results}\label{A3}

In \S 2 we evaluate the $\xi$ integrals in the above formula for $\be_{2q}(m)$ with an iterated residues technique to obtain the following simplification.

\begin{lem} \label{L2}
\begin{equation*}\be_{2q}(m) = \frac {2^{\frac {m^2+m}2}} {4\,\pi^{m}\prod_{j=1}^{m-1} j!}
\int_{Y_{2m-q} }\!\!  \ical_q(\la)\,\Delta(\lambda)\prod_{j=1}^m |\la_j|\, 
e^{-\sum_{j=1}^m \la_j}\, d\la \, ,
\end{equation*}
where
\begin{equation*}
\ical_m(\la) = \frac{2 F(D(\la))}{m+2} + \frac {16(m+2)\left(\sum_{j=1}^m \la_j\right)+16m} {m(m+1)(m+2)^3(m+3)}
-\frac{8}{(m+1)(m+2)^2} 
  \end{equation*}
and, for $m < q \leq 2m$, 
\begin{align} \label{Iq} \ical_q(\la) &= \Biggl(\frac{2 F(D(\la))}{m+2} + \frac {8 \left((m+2)^2 \la_m^2 - 2 (m+2) \la_m +2\right)}{(m+1)(m+2)^3(m+3)} \\ & \qquad \quad -  \frac{4((m+2)\la_m-1)}{(m+2)^2} \left( \frac {4\sum_{j=1}^m \la_j} {m(m+1)(m+3)}
-\frac{2}{m+1}\right) \Biggr) e^{(m+2)\la_m}\nonumber   .  
\end{align}
\end{lem}

We then use an extension of the exponential Selberg integral to calculate the exact formula for the $q=m$ case in \S 3.

\begin{lem}\label{L3}
For $m \geq 1$ and $q=m$, $$\be_{2m}(m) = \frac{4 \, m!}{\pi^m (m+2)^3(m+3)}.$$ 
\end{lem}

Unfortunately, due to the presence of the additional exponential term in $\ical_q(\la)$ when $m < q \leq 2m$, direct application of a Selberg integral formula is not possible for the other cases.  Although it would be possible to calculate the exact formulas for one or two more cases, these would be extremely complicated and we would be no closer to our goal of showing that $\be_2>0$ for all $m$.  Instead we consider the sum over the other indices and denote 
$$\be_2'(m):=\sum_{q=m+1}^{2m}\be_{2q}(m).$$ 

From Lemma \ref{L2} we immediately see that
\begin{equation*}\be_{2}'(m) = \frac {2^{\frac {m^2+m}2}} {4\,\pi^{m}\prod_{j=1}^{m-1} j!} \left( \sum_{q=m+1}^{2m}
\int_{Y_{2m-q} }\!\! \ical_q(\la)\, \Delta(\lambda)\prod_{j=1}^m |\la_j|\, 
e^{-\sum_{j=1}^m \la_j}\,  d\la \right) \, ,
\end{equation*}
where $\ical_q(\la)$ is given by \eqref{Iq}.  In \S 4 we use a change of variable argument and then exploit the symmetries in the resulting integrand to show that:
\begin{lem} \label{L2a} 
\begin{equation} \label{f1}\be_{2}'(1) =  32 c 
\int_{\R_+ }\!\!   \la_1\, 
e^{-2\la_1} d\la\, ,
\end{equation}

\begin{equation} \label{f2} \be_{2}'(2) =  48 c 
\int_{\R_+^2 }\!\! \left(\lambda _1^2 - 6  \lambda _1 +7    \right) \la_1\la_2 |(\la_1\!-\!\la_2)|\, 
e^{-\!\la_1\!-\!2\la_2}  d\la\, ,
\end{equation}
and for $m \geq 3$,
\begin{multline}\label{g} \be_{2}'(m) =  c 
\int_{\R_+^m }\!\!  d\la\, |\Delta(\lambda)|\prod_{j=1}^m \la_j\, 
e^{-\sum_{j=1}^m \la_j}\,e^{-\la_m} \Biggl(m (m+1) (m (m+3) (m+4)-4)  \\ +4(m-1) (m+2) \left( - (m+1)  (m+4)  \lambda _1  +(m-2) \la_1 \la_2 + 3 \lambda _1^2 \right) \Biggr) \, ,
\end{multline}
where
\begin{equation*}
c=\frac {2^{\frac {m^2+m-2}2}} {\,\pi^{m} (m+1) (m+2)^3 (m+3) \prod_{j=1}^{m} j!} \, .\end{equation*}
\end{lem} 
In \S 5, we prove our main result by showing that the above integrals are positive for all $m$.
\begin{theo}\label{T1}
The constant $\be_{2}(m) = \sum_{q=m}^{2m} \be_{2q}(m)$ is strictly positive for $m \geq 1$.
\end{theo}

We would like to point out that the positivity of $\be_{2q}$, for each $q$, was also conjectured in \cite{DSZ2}.  Although it seems almost certain that this is true, this remains an open problem.

Using the above result we can extend Theorem 1.6 and Corollary 1.7 from \cite{DSZ2} in the $\ncal^\crit_{N,h}$ case to all dimensions.  
We let $P(M, L)$ denote the class of positively curved metrics, i.e. metrics for which $\frac{i}{2} \Theta_h$ is a positive
(1,1)-form.  As was noted in \cite{DSZ2}, we would not expect 
$\ncal_{N,h}^\crit$ to have an upper bound as $h$ varies over $P(M, L)$, however it is bounded from below by $|c_m(L \otimes T^{* 1,0})|$.  It is therefore of interest to determine when a metric which minimizes $\ncal_{N,h}^\crit$, at least in an asymptotic sense, exists.  To this end, we make the definition:

\begin{defin} Let $L \to M$ be an ample holomorphic line bundle
over a compact \kahler manifold.   For $h \in P(M,L)$,
we say that
$\ncal_{N,h}^\crit$ is {\it
asymptotically minimal\/} if for all $ h_1
\neq h$ in $P(M,L)$, there exists $ N_0=N_0(h_1)$ such that
\begin{equation}\ncal^\crit_{N,h} \le \ncal^\crit_{N,h_1}\qquad
\mbox{for }\  N \geq N_0 \; .\end{equation}
\end{defin}

As we noted before, the first non-topological term in the asymptotic expansion of $\ncal_{N,h}^\crit$ is $\be_{2}\int_M\rho_h^2d\vol_h$.  Thus, to find minimizers of $\ncal_{N,h}^\crit$ we need to find minimizers of the Calabi functional.
We note that by results in \cite{Ca1, Ca2, Hw} all critical points of the Calabi functional are local minima.  Those that obtain the absolute minimum in a fixed \kahler class are called Calabi extremal metrics.  Therefore, the following theorem is an immediate consequence of Theorem \ref{T1}.

\begin{theo} \label{DSZTheo3}  Let $h
\in P(M,L)$ and let $\om_h=\frac i2\Theta_h$, then $\ncal_{N,h}^\crit$ is asymptotically minimal whenever $\omega_h$ is an extremal Calabi metric.
\end{theo}

By a result of Calabi in \cite{Ca2}, metrics with constant scalar curvature are extremal, and by a result of Donaldson in \cite{Don}, there is at
most one \kahler metric of constant scalar curvature in the
cohomology class $2 \pi\, c_1(L)$. Therefore, we have the corollary:

\begin{cor} \label{DSZTheo4} Suppose that $L$ possesses a metric $h$ for which the scalar curvature of
$\om_h=\frac i2\Theta_h$ is constant, then $h$ is the unique
metric on $L$ such that $\ncal_{N,h}^\crit$ is asymptotically
minimal.\label{calabi}\end{cor}

\smallskip

Due to the technical nature of this paper and the many tedious calculations in the proofs below, we have provided a Mathematica worksheet with the arXiv.org posting that provides verification in low dimensions of various intermediate steps.  It calculates the values of $\be_{2q}$ and $\be_2'$ in low dimensions using the original formula from \cite{DSZ2} and compares these with the values it calculates from the formulas given in Lemma \ref{L2}, Lemma \ref{L3}, Lemma \ref{L2a}, \eqref{B'22}, and \eqref{B'2m}.

\smallskip

These results are part of the author's ongoing thesis research at the Johns Hopkins University which is being advised by Steve Zelditch.

\section{Proof of Lemma \ref{L2} } \label{B1}

The proof of this lemma will consist of evaluating the inner integrals in \eqref{A} using an iterated residues technique.  
To this end, we rewrite $\be_{2q}(m)$ as
\begin{align*}\be_{2q}(m) &= \frac {(-i)^{m(m-1)/2}} {4\,\pi^{2m}\prod_{j=1}^{m-1} j!}
\int_{Y_{2m-q} }d\la \, \prod_{j=1}^m |\la_j|\,\Delta(\lambda)\,
\\ &\quad \times \left(F(D(\la)) \ical_{\lambda,1} + \left[ \frac {4\sum_{j=1}^m \la_j} {m(m+1)(m+3)}
-\frac{2}{m+1}\right]  \ical_{\lambda,2} + \frac {2 \ical_{\lambda,3}}{(m+1)(m+3)} \right) \;, \nonumber \end{align*} where
$$\ical_{\la,s}=\int_{\R^m}\frac {
 \Delta(\xi)\,e^{i\langle \lambda,\xi\rangle}\,d\xi}{\left(1 - \frac{i}{2}\sum\xi_j\right)^s \prod_{j\le k}\left[ 1 +\frac i2(\xi_j+\xi_k)\right]}.$$  Then we make the
change of variables $\xi_j \to t_j+i$ to obtain
$$
\ical_{\la,s} =(-1)^{\frac {m^2+m}2}(2i)^{\frac {m^2+m+2s}2}\, 
e^{-\sum \la_j}\,\ical_{\la,s,m+2}\,,$$
where
\begin{equation}\label{Aa} \ical_{\la,s,c} =
\int_{(\R-i)^m}\frac {
\Delta(t)\,e^{i\langle \lambda,t\rangle}}
 {\left(\sum t_j+ic\right)^s \prod_{1\le j \le
k\le m} (t_j+t_k)}\,dt\,. \end{equation}
Putting this together we have
\begin{align}\be_{2q}(m) &= \frac {2^{\frac {m^2+m}2}} {4\,\pi^{2m}\prod_{j=1}^{m-1} j!}
\int_{Y_{2m-q} }\,  d\la \prod_{j=1}^m |\la_j|\,\Delta(\lambda)\, 
e^{-\sum_{j=1}^m \la_j} \nonumber
\\ & \phantom{\frac {2^{\frac {m^2+m}2}} {4\,\pi^{2m}\prod_{j=1}^{m-1} j!}
\int_{Y_{2m-q} }\,} \times \Biggl(F(D(\la)) \frac {2\ical_{\lambda,1,m+2}}{i^{m^2-1}}+ \frac {2^4 \ical_{\lambda,3,m+2}}{i^{m^2-3}(m+1)(m+3)} \label{B} \\ &  \phantom{\frac {2^{\frac {m^2+m}2}} {4\,\pi^{2m}\prod_{j=1}^{m-1} j!}
\int_{Y_{2m-q} }\,} \qquad + \left( \frac {4\sum_{j=1}^m \la_j} {m(m+1)(m+3)}
-\frac{2}{m+1}\right)  \frac{2^2\ical_{\lambda,2,m+2}}{i^{m^2-2}}  \Biggr)\;. \nonumber \end{align}

To complete the proof we apply the following lemma with $p=2m-q$ and $c=m+2$ to \eqref{B} and simplify to obtain the statement of Lemma \ref{L2}.

\begin{lem} \label{LA1} Let $0\le p\le m$, $c>0$, and $s \in \{1,2,3\}$, then for
$$\la_1> \cdots >\lambda_p
> 0 > \lambda_{p+1}>\cdots
>\la_m\;,$$
we have
$$
 \ical_{\la,s,c} =\left\{\begin{array}{ll}\displaystyle
i^{m^2-s}\,\frac{\pi^m}{c^s}\, f_s(\la_m) \, e^{c\lambda_m} \quad &\mbox{for
}\ p<m\\[10pt]
\displaystyle i^{m^2-s}\,\frac{\pi^m}{c^s} &\mbox{for }\ p=m
\end{array}\right.\ ,$$
where
$$
 f_s(\la_m) =\left\{\begin{array}{ll}\displaystyle
1 \quad &\mbox{for
}\ s=1 \\[10pt]
\displaystyle 1-c\la_m &\mbox{for }\ s=2 \\[10pt]
\displaystyle \frac{c^2 \la_m^2 - 2 c \la_m +2}{2} &\mbox{for }\ s=3
\end{array}\right.\ .$$

\end{lem}

\begin{proof}
The case $s=1$ is essentially Lemma 4.3 from \cite{DSZ2}.  We will extend the proof of that lemma to prove the other cases.
To do so we let $s \in \{2,3\}$ and

\begin{equation*}
{\mathcal I}(\lambda,t;s,c) =
\frac {\Delta(t)\,e^{i\langle \lambda,t\rangle}}
 {\left(\sum t_j+ic\right)^s\prod_{1\le j \le
k\le m} (t_j+t_k)}\;,
\end{equation*}  so that
\begin{equation*}\ical_{\la,s,c}=\int_{(\R-i)^m}{\mathcal
I}(\lambda,t;s,c)\,dt .\end{equation*}

When $p>0$, we start by evaluating the
$t_1$ integral.  We close the contour of integration in the upper half plane and pick up the poles at
$t_1=0$, and at
$t_1=-t_j$ for
$j\ne 1$.  The pole at
$t_1=-ic-\sum_{j \ne 1}t_j$ is below the contour.

The $t_1=-t_j$ poles do not contribute to the integral.  To see why, we compute the residue at the pole $t_1=-t_2$ to obtain
\begin{multline*}
\frac{(-1)^{m-2} e^{i[(\lambda_2-\lambda_1)t_2+\la_3t_3+\cdots\la_mt_m]}
2t_2(t_2+t_3)\cdots(t_2+t_m)\Delta(t_2,\dots,t_m)} {(t_3+\cdots
+t_m+ci)^s\,2t_2(-t_2+t_3)\cdots(-t_2+t_m)\prod_{2\le j \le
 k\le m} (t_j+t_k)}\\ =\frac{
e^{i(\lambda_2-\lambda_1)t_2}}{2t_2}\,
\ical(\la_3,\dots,\la_m,t_3,\dots,t_m;s,c)\;.
\end{multline*}
It is easy to see that the integral of the above formula is zero, since to calculate the $t_2$ integral we would need to close the contour in the lower half plane and then the lone pole at $t_2=0$ would be above the contour.  By the symmetry in ${\mathcal I}(\lambda,t;s,c)$ we could have replaced $t_2$ in the above argument with any of the other $t_j$'s and obtained the same result.

This leaves only the pole at $t_1=0$, and the residue of 
${\mathcal I}(\lambda,t;s,c)$ at this pole is
\begin{equation}\label{C}\frac{(-1)^{m-1}}{2} {\mathcal
I}(\lambda_2,\ldots,\la_m,t_2,\dots,t_m;s,c)\;.\end{equation}
If we apply \eqref{C} recursively, we see that the integral with $0<p<m$ is reduced to
the case with all $\la_j\,$'s negative:
\begin{equation}\label{D}
{\mathcal I}_{\lambda,s,c} =
(-1)^{(m-1)+(m-2)+\cdots+(m-p)}(\pi i)^p\, \int_{(\R-i)^{m-p}}\!\!{\mathcal
I}(\lambda_{p+1},\ldots,\lambda_m,t_{p+1},\ldots,t_m;s,c) \, dt\; .
\end{equation}
When $p=m$, we compute $\mbox{Res}\;|_{t_{m} =0}\; \ical(\la_{m},t_{m};s,c)$ and see that
\begin{equation*}
{\mathcal I}_{\lambda,s,c} =
\frac{(-1)^{m(m-1)/2}(\pi i)^m}{(ic)^s}\; .
\end{equation*}

To calculate the integral in \eqref{D}, we start with the $t_m$ integral and close the contour in the
lower half plane, picking up the pole of order $s$ at $t_m=-ic-\sum_{
k<m}t_k$. These residues are
\begin{multline*}
\rcal(\la_{p+1},\dots,\la_{m},t_{p+1},\dots,t_{m-1};2,c) \!: = \\
\qquad \qquad \qquad \qquad \qquad \qquad\frac{ \Delta(t_{p+1},\ldots,t_{m-1}) \prod_{k< m} (ic+t_k+\sum_{l<m} t_l) e^{c\lambda_m + i\sum_{j<m} (\lambda_j-\lambda_m)t_j}}
{2(-ic\!-\!\!\sum_{l<m} t_l)\! \prod_{ j\le k < m} (t_j\!+\!t_k)\!
\prod_{k<m} (-ic\!-\!\!\sum_{l<m,\, l\ne k} t_l) } 
\\  \qquad \times
\left( i\la_m  -\sum_{k<m}\frac{1}{ic + t_k + \sum_{l<m} t_l} +\sum_{k }\frac{1}{ic + \sum_{l<m, \, l \neq k} t_l} \right)
, 
\end{multline*}
and 
\begin{multline*}
\rcal(\la_{p+1},\dots,\la_{m},t_{p+1},\dots,t_{m-1};3,c) \!: = \\
\qquad \qquad \qquad \qquad \qquad \qquad\frac{ \Delta(t_{p+1},\ldots,t_{m-1}) \prod_{k< m} (ic+t_k+\sum_{l<m} t_l) e^{c\lambda_m + i\sum_{j<m} (\lambda_j-\lambda_m)t_j}}
{2(-ic\!-\!\!\sum_{l<m} t_l)\! \prod_{ j\le k < m} (t_j\!+\!t_k)\!
\prod_{k<m} (-ic\!-\!\!\sum_{l<m,\, l\ne k} t_l) } 
\\  \qquad \qquad \qquad \qquad \qquad \qquad \times
\Biggl[
\left( i\la_m  -\sum_{k<m}\frac{1}{ic + t_k + \sum_{l<m} t_l} +\sum_{k }\frac{1}{ic + \sum_{l<m, \, l \neq k} t_l} \right)^2 \\  -\sum_{k<m}\frac{1}{(ic + t_k + \sum_{l<m} t_l)^2} +\sum_{k }\frac{1}{(ic + \sum_{l<m, \, l \neq k} t_l)^2} \Biggr], 
\end{multline*}
for $s=2$ and $s=3$, respectively.

Next, we evaluate the $t_{p+1}$ integral.  We close the contour in the upper half plane and see that all of the denominatorial factors in which the summand $ic$ appears either cancel out or give poles below the contour.  

It can then be verified by a straightforward (if somewhat tedious) calculation that the poles $t_{p+1}=-t_j$ do not contribute to the value of the integral.  Indeed, after computing the residue at the pole $t_{p+1}=-t_j$, consider the $t_j$ integral.  The coefficient of $t_j$ in the exponential will be $\la_j-\la_1$, which is always negative, and thus the contour can be closed in the lower half plane.  All of the poles will be above the contour, since all of the denominatorial factors with an $ic$ will have canceled out, and therefore the integral will be zero.

This leaves only the pole at $t_{p+1}=0$, and we calculate that
\begin{multline*}
 \mbox{Res}\;|_{t_{p+1} =0}\; \rcal(\la_{p+1},\dots,\la_{m},t_{p+1},\dots,t_{m-1};s,c) \\ = \frac {(-1)^{m-p-1}}{2}
\rcal(\la_{p+2},\dots,\la_{m},t_{p+2},\dots,t_{m-1};s,c)\end{multline*}
for $s \in \{2,3\}$.
We apply this argument recursively and then compute the residue of $\rcal(\la_{m-1},\la_{m},t_{m-1};s,c)$ at $t_{m-1}=0$ to obtain
\begin{equation*} \int_{(\R-i)^{m-p}}\!\!\!\!{\mathcal
I}(\lambda_{p+1},\ldots,\lambda_m,t_{p+1},\ldots,t_m;2,c) \, dt=
(-1)^{(m-p)(m-p-1)/2}(\pi i)^{m-p}\!\left(\frac{1\!-\!c\la_m}{(ic)^2}\right)
e^{c\lambda_m}\end{equation*}
and  
\begin{multline*} \int_{(\R-i)^{m-p}}\!\!\!\!{\mathcal
I}(\lambda_{p+1},\ldots,\lambda_m,t_{p+1},\ldots,t_m;3,c) \, dt \\ =
(-1)^{(m-p)(m-p-1)/2}(\pi i)^{m-p}\!\left(\frac{ c^2 \la_m^2 - 2   c \la_m +2 }{2(ic)^3}\right)
e^{c\lambda_m} \, ,\end{multline*}
for $s=2$ and $s=3$, respectively.  Substituting these formulas into \eqref{D} and simplifying gives the desired result.

\end{proof}

\section{Proof of Lemma \ref{L3}}\label{C1}

We see from Lemma \ref{L2} that 
\begin{multline*}
\be_{2m}(m) = \frac {2^{\frac {m^2+m-4}2}} {\pi^{m}\prod_{j=1}^{m-1} j!} \int_{Y_{m} }\! d\la \, \Delta(\lambda)\prod_{j=1}^m |\la_j|\, 
e^{-\sum_{j=1}^m \la_j} \\ \times \left( \frac{2 F(D(\la))}{m+2} + \frac {16(m+2)\sum_{j=1}^m \la_j +16m} {m(m+1)(m+2)^3(m+3)}
-\frac{8}{(m+1)(m+2)^2} \right)   \, .
\end{multline*}
Using \eqref{F(P)}, we can rewrite this as
\begin{multline*}
\be_{2m}(m) = c \int_{Y_{m} }\!\! d\la \, \Delta(\lambda)\prod_{j=1}^m |\la_j|\, 
e^{-\sum_{j=1}^m \la_j}\,   \Biggl(\! m (m+1) (m (m+3) (m+4)-4) \\ +4(m+2)\Biggl(-(m+1)(m+4)\sum_{j=1}^m \la_j+\left(\sum_{j=1}^m \la_j\right)^2\!\!+2\sum_{j=1}^m \la_j^2
 \Biggr)\Biggr) \, ,
\end{multline*}
where
$$ c = \frac {2^{\frac {m^2+m-2}2}} { m (m+1) (m+2)^3 (m+3)\pi^{m} \prod_{j=1}^{m-1} j!} \, .$$
Next, we see that making the change $\Delta(\lambda) \prod_{j=1}^m |\la_j| \to |\Delta(\lambda)| \prod_{j=1}^m \la_j$ in the integrand above does not change its value on the region over which we are integrating. 
After doing this, we notice that the integrand is now symmetric under permutations of $\la$, allowing us to take the integral over $\R_+^m$ and replace each of the sums with a multiple of one of the summands.  Thus
\begin{multline}\label{G}
\be_{2m}(m) = \frac{c}{m!} \int_{\R_+^{m} }\!\! d\la \, |\Delta(\lambda)|\prod_{j=1}^m \la_j\, 
e^{-\sum_{j=1}^m \la_j}\,   \Biggl(\! m (m+1) (m (m+3) (m+4)-4) \\ +4(m+2)\Biggl(-m(m+1)(m+4) \la_1+m(m-1)\la_1 \la_2 +3m \la_1^2
 \Biggr)\Biggr) \, .
\end{multline}
Now we need a lemma, which will be proved in the subsection below.

\begin{lem} \label{L4}
\begin{multline*} 
\int_{\R_+^{m} } \left( \prod_{i=1}^{k} \la_{i} \right)   \left(\prod_{i=1}^{\ell} \la_{j}  \right) \left|\Delta(\la)\right|^{2\ga} \prod_{j=1}^m\left(\la_j^{\al-1} e^{-\la_j} d\la_j \right) \\ = \ga^{k+\ell}  \left( \frac{1\!+\!\al}{\ga}\!+\!2m\!-\!\ell\!-\!k \right)_{k} \left(\frac{\al}{\ga}+m-\ell \right)_\ell  \ \prod _{j=0}^{m-1} \frac{\Ga(1 \! + \! \ga \! + \! j \! \ga)\Ga(\al \! + \! j \! \ga)} {\Ga(1 \! + \! \ga)}, 
\end{multline*}
where $(a)_n = a(a+1)\dots(a+n-1)$ is the rising factorial.  This is valid for integer $k,\ell,m$ with $0 \leq k \leq \ell < m$ and complex $\al$ and $\ga$ with Re $\al >0$, Re $\ga >$ -min $\left( \frac{1}{m}, \frac{\text{Re } \al}{(m-1)} \right).$
\end{lem}

We apply this lemma to \eqref{G} with $n=m$, $\al=2$, and $\ga=\frac{1}{2}$ for each of the four cases: $(k,l)=(0,0)$, $(k,l)=(0,1)$, $(k,l)=(0,2)$, and $(k,l)=(1,1)$ to obtain

\begin{multline*}
\be_{2m}(m) = \frac{c}{m!} \, \prod_{j=0}^{m-1} \frac{\Ga(\frac{3}{2}+\frac{j}{2})\Ga(2+\frac{j}{2})}{\Ga(\frac{3}{2})}   \, \Biggl(\! m (m+1) (m (m+3) (m+4)-4) \\ +4m(m+2)(m+3)\Biggl(-\frac{(m+1)(m+4)}{2} +  \frac{(m-1)(m+2)}{4}  +\frac{3(2m+4)}{4}
 \Biggr)\!\Biggr) \, .
\end{multline*}
After applying Gauss's multiplication formula and simplifying we have
\begin{equation*}
\be_{2m}(m) = \frac{m(m+1)\prod_{j=1}^{m}j! }{2^{\frac{m^2+m-6}{2}}} \,  c   \, .
\end{equation*}
Substituting in for $c$ gives the desired result.

\subsection{Proof of Lemma \ref{L4}} \label {C2}

First we let 
\begin{equation*}
\ical_{k,\ell} = \ical(\alpha,\gamma,k,\ell) = \int_{\R_+^{m} } \left( \prod_{i=1}^{k} \la_{i} \right)   \left(\prod_{i=1}^{\ell} \la_{j}  \right) \left|\Delta(\la)\right|^{2\ga} \prod_{j=1}^m\left(\la_j^{\al-1} e^{-\la_j} d\la_j \right) ,
\end{equation*}
$$ w_{\ell}(\la) = w(\la;\alpha,\gamma,\ell) = \left(\prod_{i=1}^{\ell} \la_{j}  \right) \left|\Delta(\la)\right|^{2\ga} \prod_{j=1}^m\left(\la_j^{\al-1} e^{-\la_j} \right),$$
and
$$ v(\la) = v(\la;\alpha,\gamma) = \left|\Delta(\la)\right|^{2\ga} \prod_{j=1}^m\left(\la_j^{\al-1} e^{-\la_j} \right).$$
Then we take the partial derivative with respect to $\la_1$ of the integrand in the formula for $\ical_{k,\ell}\,$, noting that by the Fundamental Theorem of Calculus this will then integrate to zero.  Thus,
\begin{align}
0 &= \int_{\R_+^m }\!  \frac{\partial}{\partial \la_1} \left(\prod_{i=1}^{k} \la_{i}  \  w_{\ell}(\la) \right) d\la \nonumber \\  &= \int_{\R_+^m }\!  \left(\frac{(\al+1)}{\la_1} -1 + 2 \ga \sum_{j=2}^m \frac{1}{\la_1-\la_j}  \right) \, \prod_{i=1}^{k} \la_{i}  \  w_{\ell}(\la) d\la \, . 
\label{G1}\end{align}
Here we used the fact that $$\frac{\partial}{\partial x}|x-y| = \frac{|x-y|}{x-y} \, .$$  Since $w_{\ell}(\la)$ is symmetric under permutations of the $\la_j$'s, when $j \leq k$, we see that $$ \int_{\R_+^m } \prod_{i=2}^{k} \la_{i}  \  w_{\ell}(\la) d\la = \int_{\R_+^m } \prod_{i=1}^{k-1} \la_{i}  \  w_{\ell}(\la) d\la = \ical_{k-1,\ell} \, .$$
Continuing to utilize the symmetry, we make the transposition $\la_1 \leftrightarrow \la_j$ for each of the terms in the sum in \eqref{G1}.  When $j \leq k \leq \ell$, we see that this term is zero since
\begin{equation} \label{G2} \int_{\R_+^m } \frac{\prod_{i=1}^{k} \la_{i}  \  w_{\ell}(\la)}{\la_1-\la_j} d\la = \int_{\R_+^m } \frac{\prod_{i=1}^{k} \la_{i}  \  w_{\ell}(\la)}{\la_j-\la_1} d\la = - \int_{\R_+^m } \frac{\prod_{i=1}^{k} \la_{i}  \  w_{\ell}(\la)}{\la_1-\la_j} d\la\, . \end{equation}
When $k<j\leq\ell$, we have
$$ \int_{\R_+^m } \frac{\prod_{i=1}^{k} \la_{i}  \  w_{\ell}(\la)}{\la_1-\la_j} d\la = \int_{\R_+^m } \frac{\la_j \, \prod_{i=2}^{k} \la_{i}  \  w_{\ell}(\la)}{\la_j-\la_1} d\la =  \int_{\R_+^m }  \left(1- \frac {\la_1} {\la_1-\la_j}\right) \prod_{i=2}^{k} \la_{i}  \  w_{\ell}(\la) d\la \, ,$$
and therefore 
$$\int_{\R_+^m } \frac{\prod_{i=1}^{k} \la_{i}  \  w_{\ell}(\la)}{\la_1-\la_j} d\la = \frac{1}{2} \int_{\R_+^m } \prod_{i=2}^{k} \la_{i}  \  w_{\ell}(\la) d\la = \frac{1}{2} \ical_{k-1,\ell} \, .$$
Finally, when $\ell<j\leq m$, 
$$\int_{\R_+^m } \frac{\prod_{i=1}^{k} \la_{i}  \  w_{\ell}(\la)}{\la_1-\la_j} d\la = \int_{\R_+^m }\!  \left(\la_1 \! + \! \frac{\la_1 \la_j}{\la_1-\la_j}  \right) \prod_{i=2}^{k} \la_{i}  \prod_{i=2}^{\ell} \la_{i} \  v(\la) d\la = \int_{\R_+^m } \prod_{i=2}^{k} \la_{i}  \  w_{\ell}(\la) d\la = \ical_{k-1,\ell} \, .$$
Here we applied the transposition $\la_1 \leftrightarrow \la_j$ to the second integral and by the reasoning in \eqref{G2} the second term in this integrand disappears.  Applying these results to \eqref{G1} we see that 
$$ \ical_{k,\ell} = \ga \left( \frac{\al +1}{\ga} +2m-\ell-k  \right) \ical_{k-1,\ell} \, .$$
We then iterate this relation over $k$ to achieve
\begin{equation}\label{G3} \ical_{k,\ell} = \ga^k \left( \frac{\al +1}{\ga} +2m-\ell-k  \right)_k \ical_{0,\ell} \, .
\end{equation}

Next, we apply the same technique to $\ical_{0,\ell}\,$, and we see that
\begin{align*}
0 &= \int_{\R_+^m }\!  \frac{\partial}{\partial \la_1} \left(\prod_{i=1}^{\ell} \la_{i}  \  v(\la) \right) d\la \nonumber \\  &= \int_{\R_+^m }\!  \left(\frac{\al}{\la_1} -1 + 2 \ga \sum_{j=2}^m \frac{1}{\la_1-\la_j}  \right) \, \prod_{i=1}^{\ell} \la_{i}  \  v(\la) d\la \, . 
\end{align*}
Using the arguments above, we have

\begin{equation*}
\int_{\R_+^m } \!  \frac{ \prod_{i=1}^\ell  \la_i \  v(\la)} {\la_1-\la_j}   \,  d\la = 
\begin{cases} 0, & \text{if } \ 2\leq j \leq \ell \\
\frac{1}{2} \ical_{0,\ell-1}, & \text{if  } \ \ell < j \leq m
\end{cases} \ .
\end{equation*}
Thus, 
$$ \ical_{0,\ell} = \ga \left( \frac{\al }{\ga} +m-\ell  \right) \ical_{0,\ell-1} \qquad \text{and so} \qquad \ical_{0,\ell} = \ga^\ell \left( \frac{\al }{\ga} +m-\ell  \right)_\ell \ical_{0,0} \, .$$
Combining this with \eqref{G3} produces the relation
$$ \ical_{k,\ell} = \ga^{k+\ell} \left( \frac{\al +1}{\ga} +2m-\ell-k  \right)_k \left( \frac{\al }{\ga} +m-\ell  \right)_\ell \ical_{0,0} \, .$$
To complete the proof, we note that $\ical_{0,0}$ is the so-called exponential Selberg integral, for which we have the following formula (see \cite{As}):

\begin{theo}\label{SelbergCor}
For any positive integer m, let 
\begin{equation*}  
\Phi(\la) \equiv \Phi(\la_1, \cdots , \la_m) = \left|\Delta(\la)\right|^{2\ga} \prod_{j=1}^m \left( \la_j^{\al-1} \, e^{-\la_j} \right) .
\end{equation*}
Then 
\begin{equation*}   
\int_{\R_+^m } \Phi(\la) d\la = \prod _{j=0}^{m-1} \frac{\Ga(1+\ga+j\ga)\Ga(\al+j\ga)} {\Ga(1+\ga)}, 
\end{equation*}
valid for complex $\al$, $\ga$ with Re $\al >0$, Re $\ga >$ -min $\left( \frac{1}{m}, \frac{Re \, \al}{(m-1)} \right)$.
\end{theo}

\section{Proof of Lemma \ref{L2a}}\label{D1}

We recall that from Lemma \ref{L2} we have 
\begin{equation*}\be_{2}'(m) = \frac {2^{\frac {m^2+m}2}} {4\,\pi^{m}\prod_{j=1}^{m-1} j!} \left( \sum_{q=m+1}^{2m}
\int_{Y_{2m-q} }\!\! \ical_q(\la)\, \Delta(\lambda)\prod_{j=1}^m |\la_j|\, 
e^{-\sum_{j=1}^m \la_j}\,  d\la \right) \, ,
\end{equation*}
where 
\begin{multline*} \ical_q(\la) = \Biggl(\! \frac{1}{m\!+\!2} +\frac
{4\left(\sum_{j=1}^m \la_j\right)^2+8\,\sum_{j=1}^m \la_j^2} {m(m\!+\!1)(m\!+\!2)^2(m\!+\!3)}
 + \frac {8 \left((m\!+\!2)^2 \la_m^2 - 2 (m\!+\!2) \la_m +2\right)}{(m\!+\!1)(m\!+\!2)^3(m\!+\!3)} \\  - \frac {4\sum_{j=1}^m \la_j }{m(m\!+\!1)(m\!+\!2)} -  \frac{4((m\!+\!2)\la_m\!-\!1)}{(m\!+\!2)^2} \left(\! \frac {4\sum_{j=1}^m \la_j} {m(m\!+\!1)(m\!+\!3)}
-\frac{2}{m\!+\!1}\!\right)\! \Biggr) e^{(m+2)\la_m}\nonumber   .  
\end{multline*}

For each $q$, we make the change of variables 
\begin{equation*}
\la_i \rightarrow 
\begin{cases} \la_i - \la_{2m-q+1}, & \text{for } \ 1\leq  i \leq 2m-q \\
\la_{i+1} - \la_{2m-q+1}, & \text{for } \ 2m-q <  i < m \\ 
- \la_{2m-q+1}, & \text{for } \  i = m
\end{cases}
\end{equation*}
in the integral over $Y_{2m-q}$ above.  This change of variables is a composition of the following two changes of variables:
$$
\la_i \rightarrow 
\begin{cases} \sum_{j=i}^{2m-q} \la_j, & \text{for } \ 1\leq  i \leq 2m-q \\
\sum_{j=2m-q+1}^{i} \la_j, & \text{for } \ 2m-q <  i \leq m 
\end{cases} \, 
$$
and
$$\la_i \rightarrow \la_i - \la_{i+1} .$$
These changes take $Y_{2m-q} \to \R_+^m$ and $\R_+^m \to Y_m$, respectively, and thus all of the integrals will now be over a common region of integration, $Y_m$.

Under this change of variables we see that $\Delta(\lambda)\prod_{j=1}^m |\la_j|$ is unchanged and that
$$\sum_{j=1}^m \la_j \  \to \  \sum_{j=1}^m \la_j -(m+1) \la_p \, ,$$
$$\left(\sum_{j=1}^m \la_j\right)^2 \  \to \  \left(\sum_{j=1}^m \la_j\right)^2 \!\! - 2(m+1)\la_p \left(\sum_{j=1}^m \la_j\right) + (m+1)^2\la_p^2 \, ,$$ and
$$ \sum_{j=1}^m \la_j^2 \  \to \  \sum_{j=1}^m \la_j^2 - 2 \la_p \left(\sum_{j=1}^m \la_j\right) + (m+1)\la_p^2 \, .
$$
Here we have let $p=2m-q+1$ to simplify the notation.
Therefore, since the absolute value of the Jacobian is 1, we have
\begin{equation}\label{H1}\be_{2}'(m) =  c 
\int_{Y_{m} }\!\!  \Delta(\lambda)\prod_{j=1}^m |\la_j|\, 
e^{-\sum_{j=1}^m \la_j}\, \sum_{p=1}^{m} P_p(\la) \, d\la\, ,
\end{equation}
where
\begin{equation*}
c=\frac {2^{\frac {m^2+m-2}2}} {\,\pi^{m}m (m+1) (m+2)^3 (m+3) \prod_{j=1}^{m-1} j!} \end{equation*}
and
\begin{align*}
P_p(\la) &= \Biggl(-4 (m+2) \lambda _p^2+4 (m+2) \left((m+1) (m+4)-2 \sum _{i=1}^m \lambda _i\right) \lambda _p  \\  & \qquad +m (m+1) (m (m+3) (m+4)-4) -4 (m+1) (m+2) (m+4) \sum _{i=1}^m \lambda _i \\ & \phantom{\qquad +m (m+1) (m (m+3) (m+4)-}  +4 (m+2) \left(\sum _{i=1}^m \lambda
   _i\right){}^2+8 (m+2) \sum _{i=1}^m \lambda _i^2 \Biggr)e^{-\la_p} \\
& = \Biggl(m (m+1) (m (m+3) (m+4)-4) \\ & \qquad -4 (m+1) (m+2) (m+4) \sum _{\genfrac{}{}{0cm}{3}{i=1}{i \neq p}}^{m} \lambda _i  +4 (m+2) \left(\sum _{\genfrac{}{}{0cm}{3}{i=1}{i \neq p}}^{m} \lambda
   _i\right)^2 \!\! +8 (m+2) \sum _{\genfrac{}{}{0cm}{3}{i=1}{i \neq p}}^{m} \lambda _i^2 \Biggr)e^{-\la_p} \, .
\end{align*}

Next we see that making the change $\Delta(\lambda) \prod_{j=1}^m |\la_j| \to |\Delta(\lambda)| \prod_{j=1}^m \la_j$ in \eqref{H1} does not change the value of the integrand on the region over which we are integrating. 
We make this change and now the integrand is symmetric under permutations of $\la$, so we can take the integral over $\R_+^m$. Thus
\begin{equation*}\be_{2}'(m) =  \frac{c}{m!}
\int_{\R_+^m }\!\!  |\Delta(\lambda)|\prod_{j=1}^m \left(\la_j\, 
e^{\la_j}\right)\, \sum_{p=1}^{m} P_p(\la) \, d\la\, .
\end{equation*}
The symmetry also allows us to replace any symmetric sum with a multiple of one of its summands.  Therefore
\begin{equation*}
\be_{2}'(m)  =  \frac{c}{(m-1)!}
\int_{\R_+^m }\!\!  |\Delta(\lambda)|\prod_{j=1}^m \left(\la_j\, 
e^{\la_j}\right)\,  P_m(\la) \, d\la \, .
\end{equation*}
To obtain the desired formulas, we see that 
\begin{equation*}
P_m(\la) = \left(m (m+1) (m (m+3) (m+4)-4)  \right)e^{-\la_1} = 32 e^{-\la_1}\, 
\end{equation*}
for $m=1$,
\begin{align*}
P_m(\la) & = \left(m (m\!+\!1) \left(m (m\!+\!3) (m\!+\!4)\!-\!4\right)\! +\!4  (m\!+\!2) \left(-(m\!+\!1) (m\!+\!4) \lambda _1 \!+\!\lambda
   _1^2\!+\!2 \lambda _1^2 \right)\right)e^{-\la_2} 
\\ & = 48 \left(\lambda _1^2 - 6 \lambda _1 +7  \right)e^{-\la_2}
\, 
\end{align*}
for $m=2$, and finally for $m\geq3$, we compute that
\begin{align*}
\be_{2}'(m) & =  \frac{c}{(m-1)!}
\int_{\R_+^m }\!\! d\la\, |\Delta(\lambda)|\prod_{j=1}^m \left(\la_j\, 
e^{\la_j}\right)\, e^{-\la_m} \Biggl(m (m+1) (m (m+3) (m+4)-4) \\ & \quad \qquad -4 (m+1) (m+2) (m+4) \sum _{i=1}^{m-1} \lambda _i   +4 (m+2) \left(\sum _{i=1}^{m-1} \lambda
   _i\right)^2+8 (m+2) \sum _{i=1}^{m-1} \lambda _i^2 \Biggr)  \\& =  \frac{c}{(m-1)!} 
\int_{\R_+^m }\!\!  d\la\, |\Delta(\lambda)|\prod_{j=1}^m \left(\la_j\, 
e^{\la_j}\right)\,e^{-\la_m} \Biggl(m (m+1) (m (m+3) (m+4)-4)  \\ & \qquad \qquad \qquad\qquad +4(m-1) (m+2) \left( - (m+1)  (m+4)  \lambda _1  +(m-2) \la_1 \la_2 + 3 \lambda _1^2 \right) \Biggr) \, 
\end{align*}
where once again we have replaced sums with multiples of one of their summands.

\section{Proof of Theorem \ref{T1}}\label{E1}

In \S \ref{C1} we showed that $\be_{2m}(m)$ was positive, so to prove this theorem we need to prove the positivity of $\be_2'(m)$.  
When $m=1$, there is nothing to prove since the integrand in \eqref{f1} is clearly positive.  For the other cases we need the following lemma which we will prove in \S \ref{E2}.
\begin{lem} \label{L5}
Let 
$\ical(\la)=|\Delta(\lambda)| \prod_{j=1}^m \left(\la_j\, 
e^{\la_j}\right)$,
then we have the following identities:
\begin{equation} \label{H}
\hspace{-2in} \int_{\R_+^m }\!\!  \la_1 \, e^{-\la_m} \ical(\la)  d\la  = \int_{\R_+^m }\!\!  \left(\frac{m+2}{2}+\frac{\la_1}{\la_1-\la_m} \right) e^{-\la_m} \ical(\la)  d\la \,,
\end{equation}
\begin{multline} \label{J}
\int_{\R_+^m }\!\!  \la_1^2 \, e^{-\la_m} \ical(\la)  d\la \\ = \int_{\R_+^m }\!\!  \left(\frac{(m+1)(m+2)}{2}+(m+1)  \frac{\la_1}{\la_1-\la_m} +\frac{\la_1^2} {\la_1-\la_m}\right) e^{-\la_m} \ical(\la)  d\la \,,
\end{multline}
\begin{multline} \label{L}
\int_{\R_+^m }\!  \frac{\la_1^2} {\la_1-\la_m} \, e^{-\la_m} \ical(\la)  d\la \\ = \int_{\R_+^m }\!\!  \left(\!  \frac{3 \la_1}{\la_1\!-\!\la_m} +\!\frac{(m-2)\la_1^2} {(\la_1\!-\!\la_2)(\la_1\!-\!\la_m)}+\!\frac{2\la_1^2 \, \de(\la_1\!-\!\la_m)} {|\la_1\!-\!\la_m|}\right) e^{-\la_m} \ical(\la)  d\la \,,
\end{multline}
where $\de(x)$ is the Dirac delta function, 
\begin{multline} \label{I}
\int_{\R_+^m }\!\!  \la_1 \la_2 \, e^{-\la_m} \ical(\la)  d\la \\ = \int_{\R_+^m }\!\!  \left(\frac{(m+1)(m+2)}{4}+\left( \frac{m+1}{2} \right) \frac{\la_1}{\la_1-\la_m} +\frac{\la_1 \la_2} {\la_1-\la_m}\right) e^{-\la_m} \ical(\la)  d\la \,,
\end{multline}
\begin{multline} \label{K}
\int_{\R_+^m }\!  \frac{\la_1 \la_2} {\la_1-\la_m} \, e^{-\la_m} \ical(\la)  d\la \\ = \int_{\R_+^m }\!\!  \left(\! \left(\! \frac{m+1}{2} \!\right)\! \frac{\la_1}{\la_1\!-\!\la_m} +\!\frac{\la_1 \la_2} {(\la_1\!-\!\la_m) (\la_2\!-\!\la_m)}\!-\!\frac{\la_1 \la_2} {(\la_1\!-\!\la_2)(\la_1\!-\!\la_m)}\right) e^{-\la_m} \ical(\la)  d\la \,,
\end{multline}
\medskip
\begin{equation} \label{K1}
\hspace{-2.5in} \int_{\R_+^m }\!  \frac{ \la_1 \la_2^2+\la_1^2 \la_2-2 \la_1 \la_2 \la_m}{\left(\la_1-\la_2\right)
   \left(\la_1-\la_m\right) \left(\la_2-\la_m\right)}  \, e^{-\la_m} \ical(\la)  d\la = 0 \, .
\end{equation}
The first three identities hold for $m \geq 2$ and the last three for $m \geq 3$.
\end{lem}

When $m=2$, we can apply \eqref{H} and \eqref{J} to \eqref{f2} to obtain
\begin{equation*} \be_{2}'(2) =  48 c 
\int_{\R_+^2 }\!\! \left(\frac{\la_1^2} {\la_1-\la_2}-3  \frac{\la_1}{\la_1-\la_2} +1\right) \la_1\la_2 |(\la_1\!-\!\la_2)|\, 
e^{-\!\la_1\!-\!2\la_2}  d\la\, .
\end{equation*}
Applying \eqref{L} to this gives
\begin{align} \label{B'22} \be_{2}'(2) & =  48 c 
\int_{\R_+^2 }\!\! \left(2\la_1^2 \, \de(\la_1\!-\!\la_2) +1\right) \la_1\la_2 \, 
e^{-\!\la_1\!-\!2\la_2}  d\la
\, .
\end{align}

When $m \geq 3$, we apply \eqref{H}, \eqref{J}, and \eqref{I} to \eqref{g} to obtain
\begin{multline*}\be_{2}'(m) =  c 
\int_{\R_+^m }\!\!  d\la\, |\Delta(\lambda)|\prod_{j=1}^m \left(\la_j\, 
e^{\la_j}\right)\,e^{-\la_m} \Biggl(16 (m+1)\\+2 (m-1) (m+2)\left(\frac{ -(m+1) (m+4) \lambda _1+2
   (m-2) \lambda _1 \lambda _2 + 6 \lambda _1^2}{\lambda _1-\lambda _m} \right) \Biggr).
\end{multline*}
We then apply \eqref{K} and \eqref{L} to this and we have
\begin{multline*}\be_{2}'(m) =  c 
\int_{\R_+^m }\!\!  d\la\, |\Delta(\lambda)|\prod_{j=1}^m \left(\la_j\, 
e^{\la_j}\right)\,e^{-\la_m} \Biggl(16 (m+1)\\+4 (m-1) (m+2) \Biggl( \frac{ (m-2)  \left(\la_1 \la_2^2+\la_1^2 \la_2-2 \la_1 \la_2 \la_m\right)}{\left(\la_1-\la_2\right)
   \left(\la_1-\la_m\right) \left(\la_2-\la_m\right)} + \!\frac{6 \la_1^2 \, \de(\la_1\!-\!\la_m)} {|\la_1\!-\!\la_m|} \Biggr) \Biggr).
\end{multline*}
By \eqref{K1} the middle term vanishes and therefore
\begin{equation}\label{B'2m} \be_{2}'(m) =  c\! 
\int_{\R_+^m }\!\!  d\la\, |\Delta(\lambda)|\prod_{j=1}^m \left(\la_j\, 
e^{\la_j}\right)\,e^{-\la_m} \Biggl(\!16 (m\!+\!1)+24 (m\!-\!1) (m\!+\!2)  \!\frac{\la_1^2 \, \de(\la_1\!-\!\la_m)} {|\la_1\!-\!\la_m|} \Biggr) .
\end{equation}
It is now clear that $\be_{2}'(m)$ is positive.  Indeed, by computing the $\la_1$ integral we see that
\begin{multline*} 
\int_{\R_+^m }\!\!  \!\frac{\la_1^2 \, \de(\la_1\!-\!\la_m)} {|\la_1\!-\!\la_m|} |\Delta(\lambda)|\prod_{j=1}^m \left(\la_j\, 
e^{\la_j}\right)\,e^{-\la_m} d\la \\ = \int_{\R_+^{m-1} }\!\!  \la_m^3 \, |\Delta(\la_2, \dots , \la_{m-1})|\prod_{j=2}^m \la_j\, \prod_{j=2}^{m-1} (\la_m - \la_j)^2\,
e^{-\sum_{j=2}^m \la_j}\,e^{-2 \la_m} d\la_2 \dots d\la_m .
\end{multline*}

\subsection{Proof of Lemma \ref{L5}}\label{E2}

For each of these identities the proof will consist of taking a partial derivative with respect to some $\la_i$ inside the integral on the LHS of the equation.  By the Fundamental Theorem of Calculus this will integrate to zero.  We compute the derivative and manipulate the result utilizing the symmetry in $\ical(\la)$ to achieve the desired formula.  
We will also make use of the fact that $$\frac{\partial}{\partial x}|x-y| = \frac{|x-y|}{x-y} \qquad \text{and} \qquad \frac{\partial^2}{\partial x^2}|x-y| = 2 \, \de \, (x-y).$$

So, for \eqref{H} we have
\begin{align}
0 &= \int_{\R_+^m }\!  \frac{\partial}{\partial \la_1} \left(\la_1 \, e^{-\la_m} \ical(\la)  \right) d\la \nonumber \\  &= \int_{\R_+^m }\!  \left(2 -\la_1 + \sum_{i=2}^m \frac{\la_1}{\la_1-\la_i}  \right) \, e^{-\la_m} \ical(\la) d\la \, . 
\label{M} \end{align}
Since $\ical(\la)$ is symmetric under permutations of $\la$, for $1<i < m$, we make the transposition $\la_1 \leftrightarrow \la_i$ and see that 
$$\int_{\R_+^m } \frac{\la_1 \, e^{-\la_m} \ical(\la)d\la} { \la_1-\la_i} = \int_{\R_+^m } \frac{\la_i \, e^{-\la_m} \ical(\la)d\la} { \la_i-\la_1} = 
\int_{\R_+^m} \left(1- \frac {\la_1} {\la_1-\la_i}\right) e^{-\la_m} \ical(\la)d\la,$$
and therefore 
\begin{equation} \label{N}
\int_{\R_+^m } \frac{\la_1 \, e^{-\la_m} \ical(\la)d\la} { \la_1-\la_i} = \frac{1}{2} \int_{\R_+^m } e^{-\la_m} \ical(\la)d\la.
\end{equation}
Combining \eqref{M} and \eqref{N} gives the desired result.

Next we have
\begin{align}
0 &= \int_{\R_+^m }\!  \frac{\partial}{\partial \la_1} \left(\la_1^2 \, e^{-\la_m} \ical(\la)  \right) d\la \nonumber \\  &= \int_{\R_+^m }\!  \left(3\la_1 -\la_1^2 + \sum_{i=2}^m \frac{\la_1^2}{\la_1-\la_i}  \right) \, e^{-\la_m} \ical(\la) d\la  \nonumber \\  &= \int_{\R_+^m }\!  \left((m+1)\la_1 -\la_1^2 + \frac{\la_1^2}{\la_1-\la_m}  \right) \, e^{-\la_m} \ical(\la) d\la \label{P} \, .
\end{align}
In the last equality we used the fact that, for $1<i < m$,
$$ \int_{\R_+^m }\!  \left( \frac{\la_1^2}{\la_1-\la_i}  \right) \, e^{-\la_m} \ical(\la) d\la = \int_{\R_+^m }\!  \left(\la_1+ \frac{\la_1 \la_i}{\la_1-\la_i}  \right) \, e^{-\la_m} \ical(\la) d\la = \int_{\R_+^m }\!  \la_1 \, e^{-\la_m} \ical(\la) d\la \, . $$
Here the second term in the second integral vanishes since the transposition $\la_1 \leftrightarrow \la_i$ just changes the sign of the integrand.  We then apply \eqref{H} to \eqref{P} to obtain \eqref{J}.

To obtain \eqref{L}, we see that
\begin{align*}
0 &= \int_{\R_+^m }\!  \frac{\partial}{\partial \la_1} \left(\frac{\la_1^2} {\la_1-\la_m} \, e^{-\la_m} \ical(\la)  \right) d\la \nonumber \\  &= \int_{\R_+^m }\!  \left(\frac{3 \la_1} {\la_1\!-\!\la_m} -\frac{\la_1^2} {\la_1\!-\!\la_m} +\! \sum_{i=2}^{m-1} \frac{\la_1^2} {(\la_1\!-\!\la_i)(\la_1\!-\!\la_m)}  +\!\frac{2 \la_1^2 \,  \de(\la_1\!-\!\la_m)} {|\la_1\!-\!\la_m|} \right) \, e^{-\la_m} \ical(\la) d\la  \nonumber \\  &= \int_{\R_+^m }\!  \left(\frac{3 \la_1} {\la_1\!-\!\la_m} -\frac{\la_1^2} {\la_1\!-\!\la_m} +\!  \frac{(m\!-\!2)\la_1^2} {(\la_1\!-\!\la_2)(\la_1\!-\!\la_m)}  +\!\frac{2 \la_1^2 \,  \de(\la_1\!-\!\la_m)} {|\la_1\!-\!\la_m|} \right) \, e^{-\la_m} \ical(\la) d\la \,.
\end{align*}
Here we used the fact that for $2<i<m$, the symmetry in $\ical(\la)$ implies that 
$$\int_{\R_+^m }\!  \frac{\la_1^2} {(\la_1\!-\!\la_2)(\la_1\!-\!\la_m)}  \, e^{-\la_m} \ical(\la) d\la = \int_{\R_+^m }\!  \frac{\la_1^2} {(\la_1\!-\!\la_i)(\la_1\!-\!\la_m)}  \, e^{-\la_m} \ical(\la) d\la \, .$$

For the fourth identity we have
\begin{align}
0 &= \int_{\R_+^m }\!  \frac{\partial}{\partial \la_1} \left(\la_1 \la_2 \, e^{-\la_m} \ical(\la)  \right) d\la \nonumber \\  &= \int_{\R_+^m }\!  \left(2\la_2 -\la_1 \la_2 + \sum_{i=2}^m \frac{\la_1 \la_2}{\la_1-\la_i}  \right) \, e^{-\la_m} \ical(\la) d\la \nonumber \\  &= \int_{\R_+^m }\!  \left(\frac{m+1}{2} \la_1 -\la_1 \la_2 +  \frac{\la_1 \la_2}{\la_1-\la_m}  \right) \, e^{-\la_m} \ical(\la) d\la \, . \label{O}
 \end{align}
This time we applied \eqref{N} and used the following facts which follow from the symmetry in $\ical(\la)$ as was demonstrated above: $$\int_{\R_+^m } \la_2 \, e^{-\la_m} \ical(\la)d\la = \int_{\R_+^m } \la_1 \, e^{-\la_m} \ical(\la)d\la $$
and
$$ \int_{\R_+^m } \frac{\la_1 \la_2 \, e^{-\la_m} \ical(\la)d\la} { \la_1-\la_2} = 0 \, .$$
Once again we apply \eqref{H} to \eqref{O} to obtain \eqref{I}.

To prove \eqref{K} we see that
\begin{align*}
0 &= \int_{\R_+^m }\!  \frac{\partial}{\partial \la_2} \left(\frac{\la_1 \la_2} {\la_1-\la_m} \, e^{-\la_m} \ical(\la)  \right) d\la \nonumber \\  &= \int_{\R_+^m }\!  \left(\frac{2 \la_1} {\la_1\!-\!\la_m} -\frac{\la_1 \la_2} {\la_1\!-\!\la_m} -\frac{\la_1 \la_2} {(\la_1\!-\!\la_2)(\la_1\!-\!\la_m)}+\! \sum_{i=3}^m \frac{\la_1 \la_2} {(\la_1\!-\!\la_m)(\la_2\!-\!\la_i)}  \right) \, e^{-\la_m} \ical(\la) d\la  \nonumber \\  &= \int_{\R_+^m }\!  \left(\!\frac{ (m\!+\!1)\la_1} {2(\la_1\!-\!\la_m)} -\!\frac{\la_1 \la_2} {\la_1\!-\!\la_m} -\!\frac{\la_1 \la_2} {(\la_1\!-\!\la_2)(\la_1\!-\!\la_m)}+\! \frac{\la_1 \la_2} {(\la_1\!-\!\la_m)(\la_2\!-\!\la_m)}  \right) \, e^{-\la_m} \ical(\la) d\la ,
\end{align*}
where we applied \eqref{N} to obtain the last equality.

Finally, for \eqref{K1}, we first see that
\begin{equation*} 
\int_{\R_+^m }\!  \frac{ \la_1 \la_2 \la_m}{\left(\la_1-\la_2\right)
   \left(\la_1-\la_m\right) \left(\la_2-\la_m\right)}  \, e^{-\la_m} \ical(\la)  d\la = 0 \, ,
\end{equation*}
since the transposition $\la_1 \leftrightarrow \la_2$ just changes the sign of the integrand.  Then we use the same transposition in the second term to obtain

\begin{equation*} 
\int_{\R_+^m }\!  \frac{ (\la_1 \la_2^2+\la_1^2 \la_2)\, e^{-\la_m} \ical(\la)  d\la}{\left(\la_1\!-\!\la_2\right)
   \left(\la_1\!-\!\la_m\right) \left(\la_2\!-\!\la_m\right)}   = \int_{\R_+^m }\!  \frac{ (\la_1 \la_2^2-\la_2^2 \la_1)\, e^{-\la_m} \ical(\la)  d\la}{\left(\la_1\!-\!\la_2\right)
   \left(\la_1\!-\!\la_m\right) \left(\la_2\!-\!\la_m\right)}   = 0 \, .
\end{equation*}

\begin{acknowledgements} I would like to thank my advisor, S. Zelditch, for his guidance and helpful suggestions.  I would also like to thank B. Shiffman for reviewing the manuscript and providing useful comments.
\end{acknowledgements}

{}

\end{document}